\documentclass[twocolumn,prl]{revtex4-1}
\usepackage{color}
\usepackage{amsmath}
\usepackage{amssymb}
\usepackage{graphicx}
\usepackage[unicode=true,pdfusetitle,
 bookmarks=true,bookmarksnumbered=false,bookmarksopen=false,
 breaklinks=false,pdfborder={0 0 0},pdfborderstyle={},backref=false,colorlinks=true]
 {hyperref}
\hypersetup{
 colorlinks,linkcolor=red,citecolor=blue}

\makeatletter

\usepackage{amsfonts}

\makeatother

\begin{document}

\title{Single-sideband microwave-to-optical conversion in high-Q ferrimagnetic microspheres}

\author{Cheng-Zhe Chai$^{1,2,\dagger}$, Zhen Shen$^{1,2,\dagger}$, Yan-Lei Zhang$^{1,2}$, Hao-Qi Zhao$^{1,2,\ddag}$,
Guang-Can Guo$^{1,2}$, Chang-Ling Zou$^{1,2}$, and Chun-Hua Dong$^{1,2,*}$}

\affiliation{$^{1}$CAS Key Laboratory of Quantum Information, University of Science
and Technology of China, Hefei 230026, P. R. China.}

\affiliation{$^{2}$CAS Center For Excellence in Quantum Information and Quantum
Physics, University of Science and Technology of China, Hefei, Anhui
230026, P. R. China.}

\thanks{$^{\dagger}$These authors contributed equally to this work}
\thanks{$^{\ddag}$Present address: Department of Electrical and Systems Engineering, University of Pennsylvania, Philadelphia, PA 19104, USA}
\thanks{$^{*}$chunhua@ustc.edu.cn}

\date{\today}

\maketitle
\textbf{Coherent conversion of microwave and optical photons can significantly expand the
ability to control the information processing and communication systems. Here, we experimentally demonstrate the microwave-to-optical
frequency conversion in a magneto-optical whispering gallery mode
microcavity. By applying a magnetic field parallel to the microsphere
equator, the intra-cavity optical field will be modulated when the
magnon is excited by the microwave drive, leading to microwave-to-optical
conversion via the magnetic Stokes and anti-Stokes scattering processes.
The observed single sideband conversion phenomenon indicates a non-trivial
optical photon-magnon interaction mechanism, which is derived from
the magnon induced both the frequency shift and modulated coupling
rate of optical modes. In addition, we demonstrate the single-sideband
frequency conversion with an ultrawide tuning range up to 2.5GHz, showing
its great potential in microwave-to-optical conversion.}

\section{introduction}

Electromagnetic waves at microwave and optical frequencies play important roles in information processing and communication systems. However, the quantum information
technology based on the most promising superconducting qubits is operated at cryogenic
temperature with microwave photons, which cannot achieve long distance
communications between qubits. Unlike microwave photons, the optical photon can transmit via low
loss optical fibers, making them suitable for long distance communication. Thus frequency conversion between microwave
photon and the optical photon has attracted great interest \cite{andrews2014bidirectional,bochmann2013nanomechanical,rueda2016efficient,lambert2020coherent,fan2018superconducting,han21}.
Besides, the energy of single microwave photon is too low to be efficiently
detected with a high signal-to-noise ratio. In contrast, converting
the microwave photons to optical photons can be detected directly
with single-photon detectors. Therefore, it can greatly promote the
detection based on the microwave, help to improve the resolution of radar,
and maybe realize the quantum enhanced radar system\textbf{~}\cite{PhysRevLett.114.080503}.
Recently, such a microwave to optical transducer~\cite{han21}
has been demonstrated in optomechanics,
electro-optic interaction,
atoms and ions~\cite{vainsencher2016bi,balram2016coherent,soltani2017efficient,higginbotham2018harnessing,xu2021bidirectional,vogt2019efficient,PhysRevA.96.013833,Petrosyan_2019,welinski2019electron,everts2019microwave}. Among
these approaches, optomagnonics based on magnon provide an alternative
and attractive approach of the coherent microwave-to-optical conversion
because of its great frequency tuning range and long coherence time~\cite{chumak2015magnon,harder2018level,Tabuchi405,Lachance-Quirion425,Lachance_Quirion_2019,PhysRevA.94.033821,graf2018cavity}.

\begin{figure}
\centerline{\includegraphics[clip,width=0.8\columnwidth]{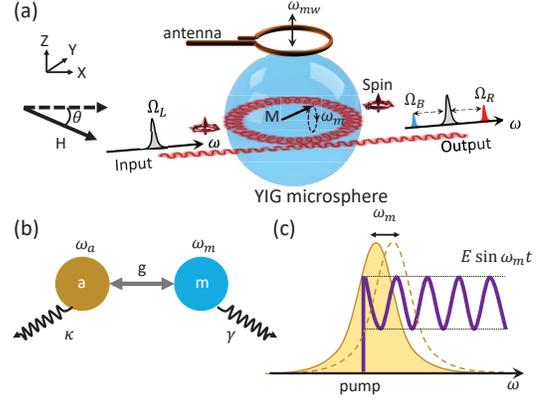}}\caption{(a) Schematic of the microwave-to-optical frequency conversion in
a YIG microsphere. The bias magnetic field is parallel to the optical
path, while the input light excites the WGMs and has an optical spin
perpendicular to the propagating direction due to the spin-orbit coupling.
The intra-cavity field could be modulated by the dynamic magnetic
field via the Faraday effect, and generate two sidebands at the output
port. (b) Frequency conversion via the coupling between photon ($a$)
and magnon ($m$). (c) Illustration of the dispersive magnon-photon
coupling as the magnetization induced modulation of the resonant frequency.}

\label{Fig1}
\end{figure}

Currently, frequency conversion has been demonstrated in such an optomagnonic
system~\cite{zhang2016optomagnonic,osada2016cavity,PhysRevLett.120.133602,PhysRevLett.117.133602,zhu2020waveguide},
where the high Q yttrium iron garnet (YIG) whispering gallery mode
(WGM) microcavity was used to enhance the interaction between magnons
and photons, and non-reciprocity of the magnetic Brillouin light scattering
(BLS) has been observed~\cite{hisatomi2019helicity,PhysRevLett.120.133602}.
However, similar to the Brillouin optomechanical system~\cite{dong2015brillouin,optomechanics},
the triple-resonance condition (the phase matching between pump, signal
and a magnetic modes) is required in such system, which may limit
the flexibility in choosing the working frequencies and tunability
of the frequencies. Therefore, using the great tunability of the magnon
and also two-mode magnon-photon coupling mechanism would allow us
to achieve transducer that mitigates the above limitations.

In this Letter, a tunable frequency conversion between microwave and
photons is realized by the dynamical Faraday effect in a YIG microsphere.
The magnetic Stokes and anti-Stokes scattering induced by the dispersive
interaction between magnon mode and optical mode can be observed.
When the frequency of pump light is resonant with optical mode, the
asymmetry of the two sidebands even single sideband (SSB) is observed
in our experiment. By changing the direction of the static external
magnetic field, we observed both the optical mode resonance frequency
shift and modulated coupling efficiency, corresponding to both phase
and intensity modulations. Therefore, we deduced that this asymmetry
conversion is derived from the phase and intensity modulations induced
by the internal magnetization procession. In our experiment, we demonstrated
16 times asymmetry of the two sidebands and the magnon tuning range
of 2.5 GHz, corresponding to the tunable frequency conversion with
same range. Our results serve as a novel method for the implementation
of SSB microwave-to-optical conversion devices.

\section{Experimental Setup and Results}
The principle of the frequency conversion in YIG microsphere is illustrated
in Fig.~\ref{Fig1}(a). The input light couples to the YIG microcavity
and excites the WGMs through the high-index prism. The WGMs in microcavity
will have a spin along the z direction due to the spin-orbit coupling
of light~\cite{chai2020non,Bliokh2015b,Bliokh2015a}. According to
our previous work, the spin will be modulated by the magnetization
along z direction due to Faraday effect, thus shift the resonant frequency
of the WGMs~\cite{chai2020non}. When applying a magnetic field parallel
to the resonator equator, the microwave excites the magnon mode in
microcavity by an antenna and causes the procession of the magnetization
in microcavity. Therefore, as shown in Fig.~\ref{Fig1}(c), the resonant
frequency of the optical mode in the microcavity will be modulated
by the magnetization procession. When an optical pump drives at the
optical mode, its amplitude will be modulated and lead to two sidebands
at the output as oscillator system, and the Hamiltonian of the system
can be written as
\begin{eqnarray}
H & = & \omega_{a}a^{\dagger}a+\omega_{m}m^{\dagger}m+ga^{\dagger}a(m+m^{\dagger}),
\end{eqnarray}
where $a$ ($a^{\dagger}$), $m$ ($m^{\dagger}$) are the annihilation
(creation) operators for optical mode and magnon mode, respectively.
$g$ is the magneto-optical coupling strength, and $\omega_{a}$ and
$\omega_{m}$ are the frequency of the optical mode and magnon mode,
respectively. Different from the previous experiment based on triple-resonance
condition \cite{zhang2016optomagnonic,osada2016cavity,PhysRevLett.120.133602,PhysRevLett.117.133602},
only one optical mode and one magnon mode are participated in the
magneto-optical interaction, as shown in Fig. \ref{Fig1}(b). Therefore,
the interaction between photon and magnon in our experiment is similar
to the two mode interaction in optomechanical system \cite{shen2016experimental}.
Considering the great tunability of the magnon, a larger operating frequency
transducer could be obtained than previous research in optomagnonic
system.

\begin{figure}
\centerline{\includegraphics[clip,width=0.8\columnwidth]{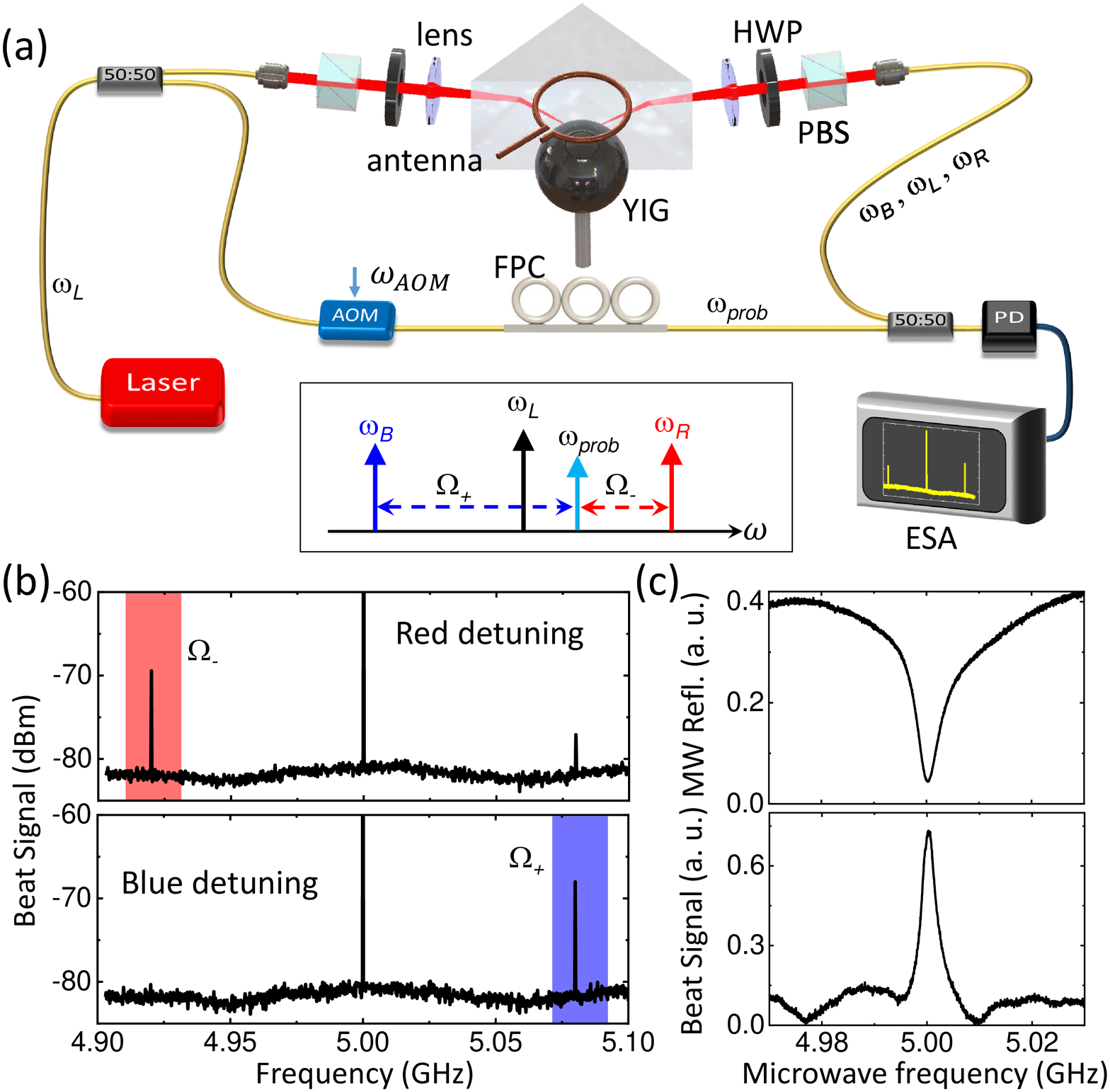}}\caption{(a) Schematic of experimental setup. A tunable laser is separated
into two beams by an optical fiber splitter. One beam excites the
WGMs by prism coupling, which would generate two sidebands at output
and then combine another beam as the local oscillator (LO) shifted
by an acousto-optic modulator (AOM). PBS: polarization beam splitter.
HWP: half wavelength plate. FPC: fiber polarization controller. PD:
photon detector. ESA: electric spectra analyzer. Inset: Spectral position
of the pump laser, the sidebands and the probe laser as the local oscillator.
(b) The detected beat signal when the optical pump at red and blue
detunings, respectively.\textcolor{red}{{} }\textcolor{black}{The $\Omega_{-}(\Omega_{+})$
corresponding to the sideband signal, which is higher (or lower) than
the optical pump.} (c) Microwave reflection and the generated beat
signal as a function of the microwave frequency.}

\label{Fig2}
\end{figure}

Figure \ref{Fig2}(a)
is the schematic of our experimental setup. A tunable diode laser
is seperated into two laser beams by an optical fiber splitter. One
beam excites the WGMs by the rutile prism, and would generate two
sidebands due to the interaction between the magnons and photons.
Another laser beam is shifted with frequency of +80$\mathrm{MHz}$
by an acousto-optic modulator (AOM) as the local oscillator (LO) to
measure the sidebands through heterodyne measurement, as shown in
the inset of Fig.~\ref{Fig2}(a). Two laser beams are combined with
the beam splitter and sent a 12$\,\mathrm{GHz}$ high-speed photon
detector followed by a microwave amplifier and a spectrum analyzer.
In our experiment, the radius of the YIG microsphere is about 400$\,\mathrm{\mu m}$.
Two sets of polarization beam splitters and half-wave plates are used
to control the polarization of the system. One is for exciting the
different polarized WGMs, and another one is for verifying that the
polarization of the generated sideband signal, which is consistent
with the pump light. And a fiber polarization controller is used to
make two light paths have the same polarization for optimizing the beat
signal. The magnon mode used in the experiment is the uniform mode
as know as Kittel mode, whose frequency is determined by $\omega_{m}=\gamma H,$
where $\gamma=2\pi\times2.8\,\mathrm{MHz/Oe}$ is the gyromagnetic
ratio and H is the external bias magnetic field. The external static
magnetic field with intensity of approximately 1780$Oe$ is parallel
to the resonator equator, corresponding to the magnon frequency of
approximately $5\,\mathrm{GHz}$, and the Kittel mode is excited by
an antenna placed above the YIG microsphere, with the microwave power
is amplified up to about $500\,\mathrm{mW}$.

The pump laser with frequency of $\omega_{L}$ modulated by the dynamic
Faraday effect through the YIG microcavity will be scattered to two
sidebands ($\omega_{R}$ and $\omega_{B}$). The LO beam has a $+80\,\mathrm{MHz}$
frequency shift with respect to the pump laser.\textcolor{red}{{} }\textcolor{black}{As
a result, the sideband signal $\omega_{R}$ (or $\omega_{B}$) is
measured through the beat signals} \textcolor{black}{$\Omega_{-}$
(or $\Omega_{+}$)} in spectrum analyzer, as shown in Fig.~\ref{Fig2}(b).
The typical results are measured when the optical pump has a red (or blue) detuning from the optical mode set equal to the magnon frequency, which indicate
the Anti-Stokes (or Stokes) scattering. Figure~\ref{Fig2}(c) shows
the microwave reflection and the generated beat signal ($\Omega_{-}$)
as a function of the microwave frequency via a vector network analyzer
(VNA). The beat signal shows a resonant characteristics and correlates
to the magnon mode which verified the participation of the magnon
in the frequency conversion process.

\begin{figure}[tp]
\centerline{\includegraphics[clip,width=0.8\columnwidth]{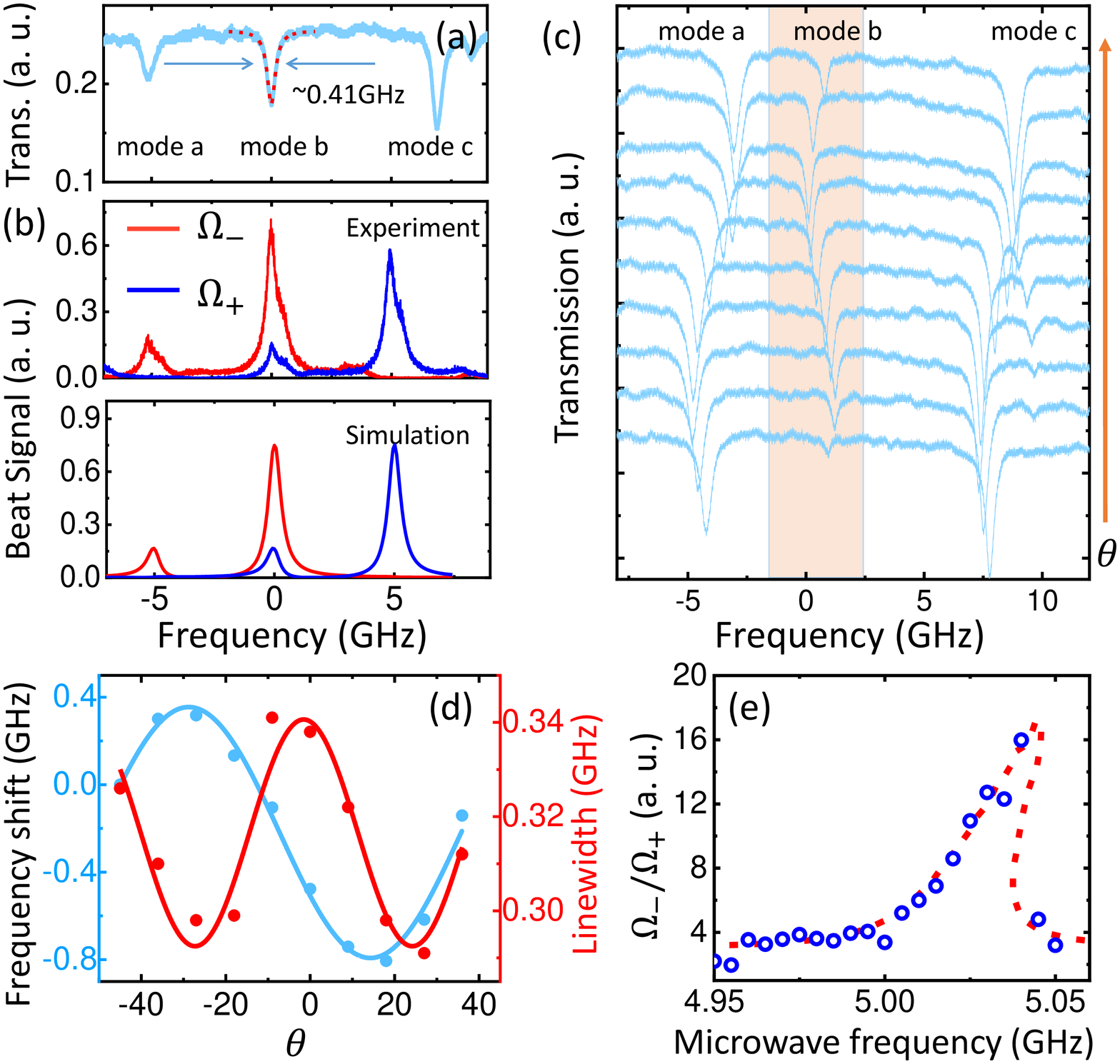}}\caption{(a-b) Optical pump transmission and the generated optical signal as
a function of pump laser frequency. The experimental results are agreed
well with the numerical calculations. (c) Transmission as a function
of the input light frequency with different bias magnetic field direction.
(d) The frequency shift and linewidth of the optical mode as a function
of the bias magnetic field direction from (c). (e) $\Omega_{-}/\Omega_{+}$
ratio as a function of the input microwave frequency, the red dotted
line represents the fitting result when considering the thermal effect
of magnon.}

\label{Fig3}
\end{figure}

When the pump laser is scanning through the cavity modes, the typical
transmission is shown in Fig. \ref{Fig3}(a). The dotted line is Lorentz
fitting, corresponding to the loaded Q factor of $4.7\times10^{5}$.
Figure \ref{Fig3}(b) shows the converted signals (\textcolor{black}{$\Omega_{-}$
or $\Omega_{+}$}) obtained with the same optical mode by scanning the pump laser. When the pump laser is red detuning with magnon
frequency, the sideband $\omega_{R}$ is resonant with the optical
mode and another sideband $\omega_{B}$ is far detuned, thus generate
the fairly strong sideband $\omega_{R}$ in optical spectrum, the
anti-Stocks scattering dominant at this situation. Therefore, the
signal of $\Omega_{-}$ is only observed. To the contrary, the signal
of $\Omega_{+}$ is only observed when the pump laser is blue detuning.
Especially, when the pump laser is resonant with optical mode, the
modulation of optical resonant would induce two sidebands, as shown
both obvious the signal of $\Omega_{-}$ and $\Omega_{+}$. Therefore,
the signal of $\Omega_{-}$ and $\Omega_{+}$ has the similar shape
just with a magnon frequency shift.

Surprisingly, when the pump laser is resonant with optical mode,
the two sidebands signals are asymmetric in Fig. \ref{Fig3}(b), in
contrast to the symmetric sidebands reported in the dispersively coupled
optomechanical systems~\cite{Kippenberg1172,Wollman2015}. To investigate
the physical mechanism for this novel SSB phenomenon, Fig.~\ref{Fig3}(c)
shows the optical transmission with changing the magnetic field direction,
corresponding to the magnetic field direction changed with the microwave
signal. The angle $\theta$ between the magnetic field direction and
the resonator equator is changed from the $-45^{\circ}$ to $40^{\circ}$.
Especially, the transmission of the target optical mode for the efficient
frequency conversion has the obvious change during the adjustment,
as shown in the gray region of Fig.~\ref{Fig3}(c), while the transmission
of other optical modes have no obvious change during the process.
These results indicate that the changed magnetization direction of
the YIG sphere could significantly change the coupling strength of the
target resonance mode with near-field coupler. Figure~\ref{Fig3}(d)
further shows the resonant frequency and linewidth as a function of
the $\theta$. It indicates that both the resonant frequency and linewidth
of the optical mode could also be modulated by the dynamic magnetic
field and the two modulations have a phase difference $\sim\pi$,
corresponding to novel magnon-photon coupling that induces out-of-phase
dispersive and dissipative modulations. Similar to the optical SSB
modulator~\cite{1291515}, such out-of-phase phase modulation (frequency)
and intensity modulation (coupling strength) eventually leads to the
SSB microwave-to-optical conversion.

To consider the optical mode owning two modulations with the dynamic
magnetic field induced by the microwave, one is frequency modulation
$ga^{\dagger}a(m+m^{\dagger})$ shown in Eq.(1), and another one is
coupling strength modulation which $\kappa_{a,1}$ could be expressed
as $\kappa_{a,1}[1+A(m+m^{\dagger})]$, where $A$ is the modulation
coefficient of the coupling strength. The coupled-oscillator equations
in interaction picture are then rewritten as
\begin{eqnarray}
\frac{da}{dt} & = & [i\Delta_{a}-\frac{\kappa_{a}}{2}]a-iga(me^{-i\omega_{mw}t}+\mathbf{\mathrm{H}}.\mathrm{c}.)\nonumber \\
 &  & +\sqrt{\kappa_{a,1}}\left[1+A\left(me^{-i\omega_{mw}t}+\mathbf{\mathrm{H}}.\mathrm{c}.\right)/2\right]\sqrt{\frac{P_{L}}{\hbar\omega_{L}}},
\end{eqnarray}

\begin{eqnarray}
\frac{dm}{dt} & = & [i\Delta_{m}-\frac{\kappa_{m}}{2}]m-iga^{\dagger}ae^{i\omega_{mw}t}+\sqrt{\kappa_{m,1}}\sqrt{\frac{P_{mw}}{\hbar\omega_{mw}}}\nonumber \\
 &  & +(A\sqrt{\kappa_{a,1}}/2)\sqrt{\frac{P_{L}}{\hbar\omega_{L}}}(a-a^{\dagger})e^{i\omega_{mw}t},
\end{eqnarray}
where $\Delta_{a}=\omega_{L}-\omega_{a}$ and $\Delta_{m}=\omega_{mw}-\omega_{m}$
are the detuning of the optical and the microwave pump, respectively.
$\kappa_{a}$, ($\kappa_{m}$) is the decay rate of the optical (magnon)
mode. The numerical results are shown in Fig. \ref{Fig3}(b). In Fig.
\ref{Fig3}(b), one can see that the experimental and numerical results
are in good agreement which further verify the two modulation on optical
mode causes the SSB phenomenon. Further studies have found that when
the pumped laser is kept at the near resonance position, the ratio
of beat frequency signal between red and blue ($\Omega_{-}/\Omega_{+}$)
sideband changes with different microwave detuning are measured. The
magnon frequency is $5\,\mathrm{GHz}$ and it is found that the ratio
of beat frequency signal has distinct resonant properties, which means
the stronger driving force of microwave enhances the precession amplitude
of the magnon along the z-axis near the resonator equator, leading
to the stronger SSB effect as shown in Fig.~\ref{Fig3}(e). The resonant
frequency in Fig. \ref{Fig3}(e) is larger than $5\,\mathrm{GHz}$
because of the thermal effect on the magnon~\cite{chaiAPL}.

\begin{figure}
\centerline{\includegraphics[clip,width=0.8\columnwidth]{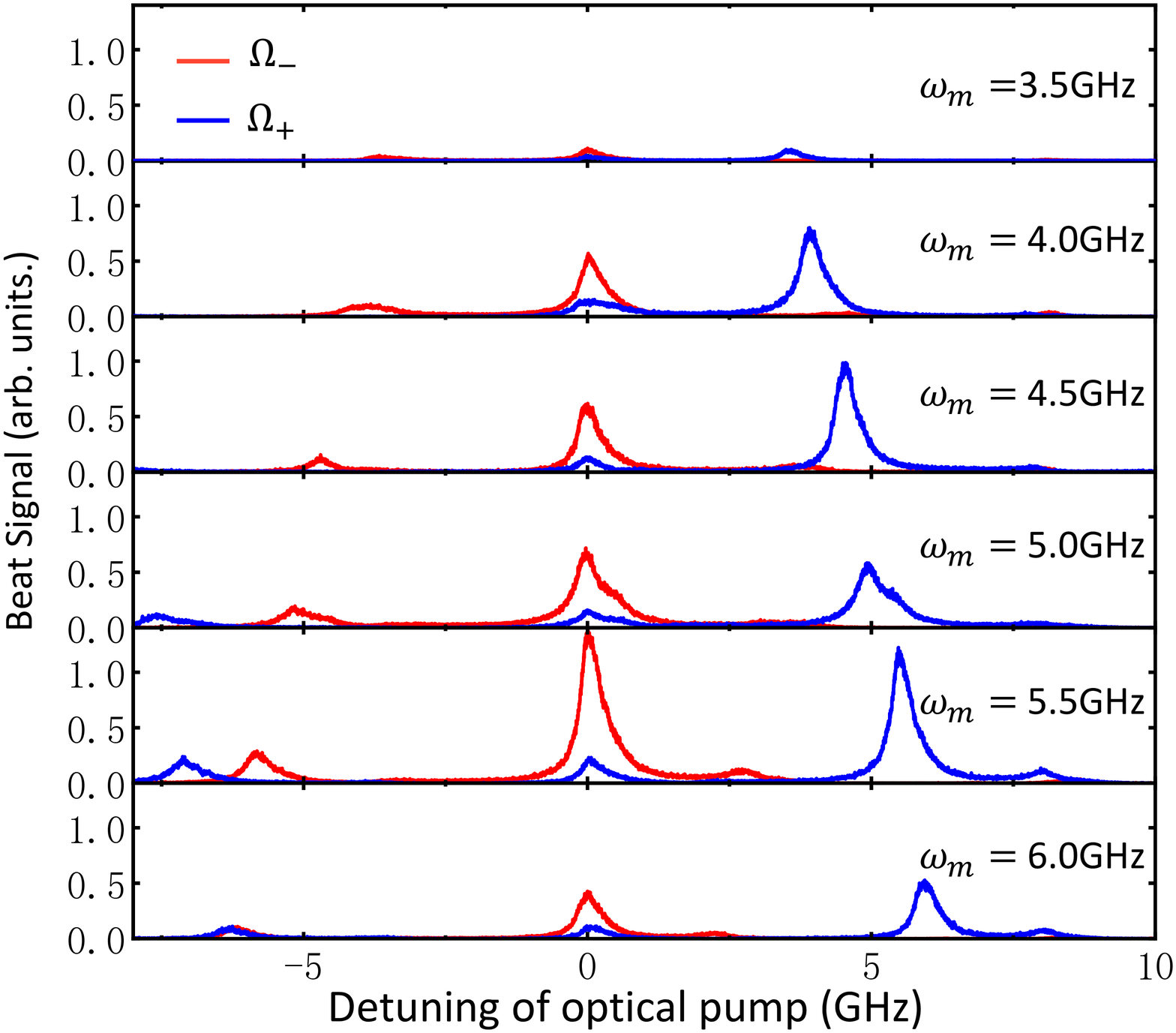}}\caption{The converted signal
as a function of pump laser frequency with different magnon frequencies by changing magnetic field.}

\label{Fig4}
\end{figure}
As we know, the external magnetic field can tune the frequency of the magnon, corresponding to the tunable frequency conversion. Figure~4 shows the frequency conversion by only tuning the relevant
magnon frequency from $3.5$ to $6\,\mathrm{GHz}$, which range is
limited by the permanent magnet we used. It shows that our device
has a much larger operation bandwidth compared to the previous scheme~\cite{zhang2016optomagnonic,PhysRevLett.117.133602,osada2016cavity,PhysRevLett.120.133602}.
Taking the insertion losses into consideration, the power conversion efficiency
in our experiment is
estimated as $3.62\times10^{-6}$. Despite the conversion efficiency is lower
than that the previous experiment based on triple-resonance condition~\cite{zhang2016optomagnonic,osada2016cavity,PhysRevLett.117.133602,zhu2020waveguide,PhysRevLett.120.133602},
our experiment provides a novel method to realize the SSB frequency conversion between microwave and optical photons. And the conversion efficiency can be improved by introducing a second microcavity to approach the
first microcavity, so that the optical mode will split due to strong
coupling between two microcavities~\cite{soltani2017efficient}. The
coupling is tunable, and so the splitting of the optical mode can
be selected to match the required magnon frequency to achieve the optical
pump and sideband signal resonant with the optical mode, allowing
for larger frequency conversion efficiency. Besides, scaling down
the size of the YIG microsphere to reduce the mode volume, or replace
the microsphere with microdisk and use high order magnetic mode to
improve the mode overlap, can further improve the conversion efficiency~\cite{zhu2020waveguide,PhysRevLett.120.133602,doi:10.1063/1.4907694,PhysRevB.97.214423}.

\section{Conclusion}
We experimentally demonstrate the SSB frequency conversion between microwave and optical photons
in a YIG microcavity. By apply the static magnetic field parallel
to the resonator equator, the magnetic Stokes and anti-Stokes scattering
occurs in YIG microsphere, which resembles the phonon-photon interaction
in an optomechanical system. Besides, due to the modulation of both
the resonance frequency and external coupling intensity of optical
modes, the SSB modulation has been demonstrated in our experiment.
Our results provide a novel method to realize the SSB frequency transducer.

\section*{Funding}
The work was supported by the National Key R\&D Program of China (Grant
No.2020YFB2205801), the National Natural Science Foundation of China
(Grant Nos.11874342, 11934012, 61805229, and 92050109), USTC Research Funds of the Double First-Class Initiative (YD2470002002) and the Fundamental
Research Funds for the Central Universities. C.-H. Dong was also supported
the State Key Laboratory of Advanced Optical Communication Systems
and Networks, Shanghai Jiao Tong University, China. This work was
partially carried out at the USTC Center for Micro and Nanoscale Research
and Fabrication.

\section*{Disclosures}
The authors declare no conflicts of interest.

\end{document}